\documentclass{iau}

\usepackage{amsmath}
\usepackage{graphicx}
\usepackage{multirow}

\begin{document}

\lefttitle{Ge \& Han}
\righttitle{Proceedings of the IAU Symposia 389: Gravitational Wave Astronomy}

\jnlPage{1}{8}
\jnlDoiYr{2024}
\doival{xx.xxxx/xxxx}

\aopheadtitle{Proceedings IAU Symposium}
\editors{David Buckley \& Paul Groot, eds.}

\title{Mass Transfer Physics in Binary Stars and Applications in Gravitational Wave Sources}

\author{Hongwei Ge \& Zhanwen Han}
\affiliation{Yunnan Observatories, Chinese Academy of Sciences,
	Kunming, 650216, P.R. China, \email{gehw@ynao.ac.cn}}
\affiliation{Key Laboratory for Structure and Evolution of Celestial Objects, 
	CAS, Kunming 650216, P.R. China}
\affiliation{International Centre of Supernovae, Yunnan Key Laboratory,
	Kunming 650216, P.R. China}

\begin{abstract}
The stability criteria of rapid mass transfer and common-envelope evolution are fundamental in binary star evolution. They determine the mass, mass ratio, and orbital distribution of many important systems, such as X-ray binaries, type Ia supernovae, and merging gravitational-wave sources. In the limit of extremely rapid mass transfer, the response of a donor star in an interacting binary becomes asymptotically one of adiabatic expansion. We built the adiabatic mass-loss model and systematically surveyed the thresholds for dynamical timescale mass transfer over the entire span of possible donor star evolutionary states. Many studies indicate that new mass transfer stability thresholds play an essential role in the formation and properties of double compact object populations and the progenitors of SNe Ia and detectable GW sources. For example, our studies show that the mass transfer in the red giant and the asymptotic giant branch stars and the massive stars can be more stable than previously believed. Consequently, detailed binary population synthesis studies, using updated unstable mass transfer criteria, predicate the non-conservative stable mass transfer may dominate the formation channel of double stellar-mass black holes and can explain the population of the large mass ratio double stellar-mass black holes. Using our updated mass transfer thresholds, binary population thesis studies by Li et al. show that Ge et al.'s results support the observational double white dwarfs merger rate distribution per Galaxy and the space density of double white dwarfs in the Galaxy.
\end{abstract}

\begin{keywords}
Gravitational Waves, Binary Stars, White Dwarfs, Neutron Stars, Black Holes
\end{keywords}

\maketitle

\section{Introduction}
\begin{figure}
	\includegraphics[scale=.09]{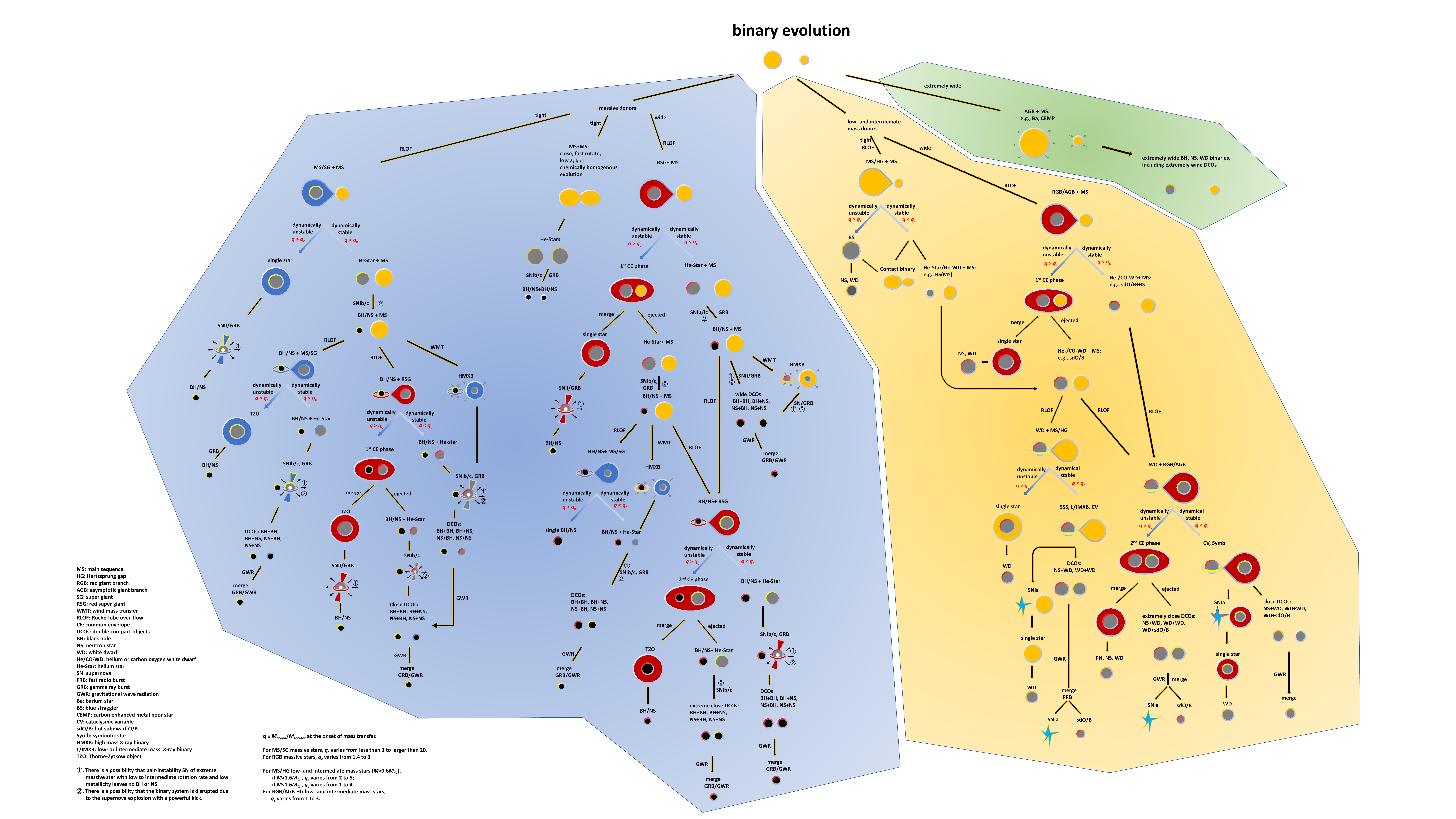}
	\caption{Binary star evolution tree from \cite{2020RAA....20..161H}.}
	\label{fig01}
\end{figure}

Over the past two decades, many fundamental discoveries have improved our understanding of modern astrophysics. For instance, people use the type Ia supernovae as distance indicators and find that the universe's expansion accelerates, providing the quantitative confirmation of the dark energy \citep{1998AJ....116.1009R,1999ApJ...517..565P,2003ApJ...594....1T,2004ApJ...607..665R}. The ground-base LIGO-Virgo detectors opens a new window for gravitational wave (GW) astronomy, of which brings binary stellar physics a new era and urges our improved knowledge about the formation and evolution of stellar-mass black hole (smBH) binaries \citep{2016PhRvL.116f1102A,2020PhRvL.125j1102A,2020ApJ...896L..44A}.
Four worldwide Pulsar Timing Array collaborations announce the detection of nanohertz GWs by inspiring supermassive black hole binaries using millisecond pulsars \citep{2023RAA....23g5024X,2023ApJ...951L...8A,2023ApJ...951L...6R,2023A&A...678A..50E}. Binary stars' evolution form not only type Ia supernovae, smBH binaries, and millisecond pulsars mentioned above but also almost all kinds of exotic binaries (Figure\,\ref{fig01}) such as crucial space LISA \citep{2023LRR....26....2A}/Tianqin \citep{2024arXiv240919665L} GW sources double white dwarfs (DWDs, \citealt{2023A&A...669A..82L}), high/low-mass X-ray binaries, cataclysmic variables, blue stragglers, hot subdwarfs \citep[][and references therein]{2015ApJ...812...40G,2020ApJ...899..132G,2020ApJS..249....9G,2020RAA....20..161H}.

\begin{figure}
	\includegraphics[scale=0.45]{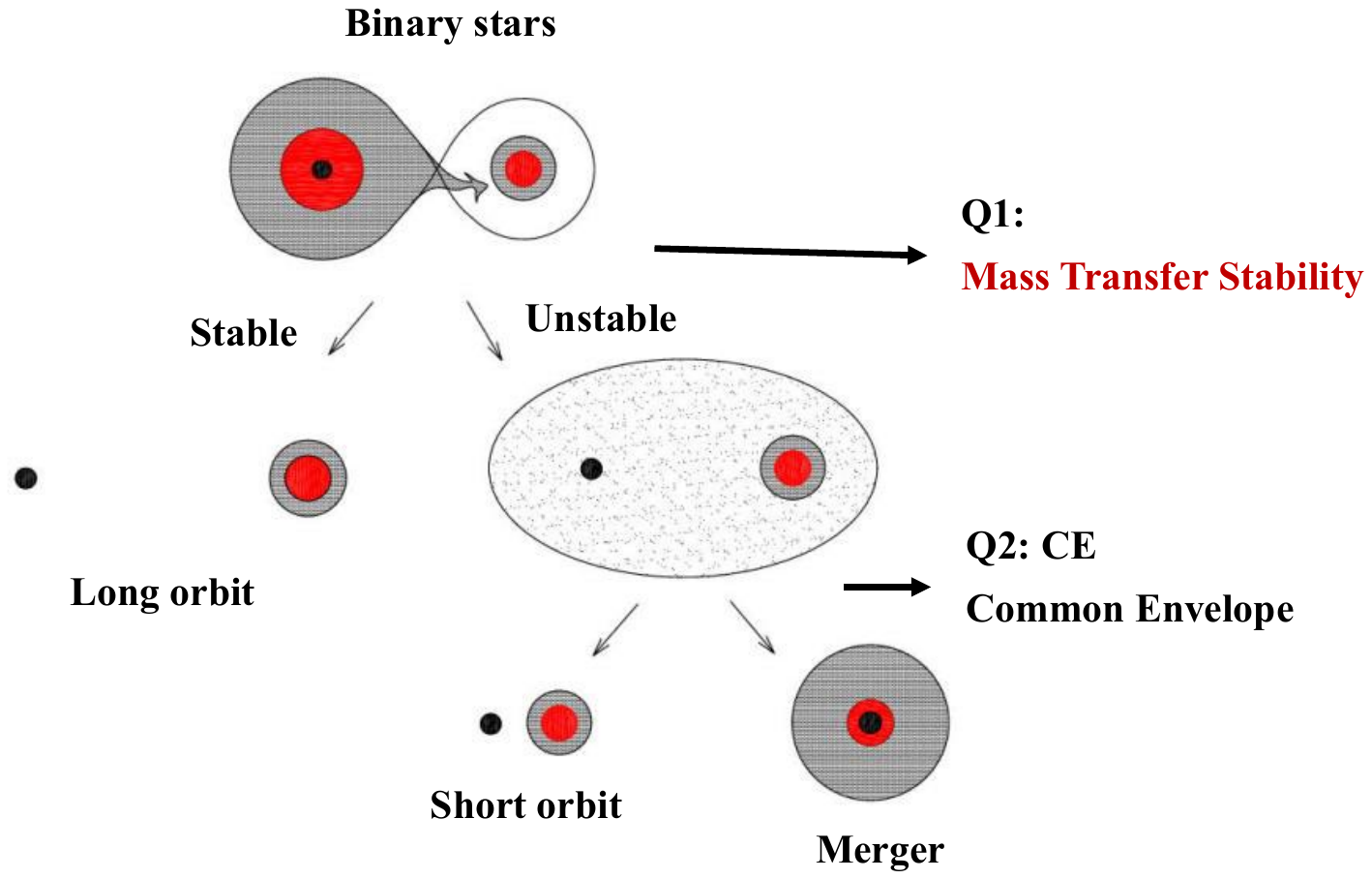}
	\caption{Two fundamental questions in binary star evolution modified and adopted from \cite{2003ASPC..289..413H}. The accretor can be a normal or compact star, and multiple binary mass transfer processes can happen during binary star evolution. }
	\label{fig02}
\end{figure}

Despite the importance of binary star evolution, two fundamental questions (Figure\,\ref{fig02}), i.e., the criteria of unstable mass transfer \citep{1987ApJ...318..794H,1997A&A...327..620S} and the common envelope evolution \citep[CEE;][]{1976IAUS...73...75P,1975PhDT.......165W}, have yet to be better understood over the last few decades. For example, the merger rate of double compact binaries can vary for orders of magnitudes without the precise constraint CEE \citep{2002ApJ...572..407B}. The intermediate/old populations of type Ia supernovae arising from the symbiotic channel under a revised mass-transfer prescription can be significantly enlarged \citep{2019A&A...622A..35L}. These two fundamental questions in binary star evolution determine the evolution fate of binary stars, such as the formation channels, rates, mass, and orbital period distributions of many binary objects. They are also critical physical inputs for binary population synthesis codes in studying the formation and evolution of large samples of binaries.

In the next few decades, a combination of full or multiple wavelength electromagnetic observations with ground-based LIGO/Virgo/KAGRA collaborations and space LISA/Tianqin low-frequency GW detectors will provide mega-data compact objects containing BHs, neutron stars (NSs), or WDs. This inspires us to precisely constrain the fundamental questions in binary star evolution.

\section{Progress in mass transfer stability thresholds}

Significant progress has been made in the criteria of unstable mass transfer since four decades ago. \citet{1987ApJ...318..794H} used simplified polytropic stellar models undergoing adiabatic mass loss for investigating the mass transfer stability. \citet{1997A&A...327..620S} further extended the study to non-conservative mass transfer cases. People also used detailed stellar evolution codes such as STARS and MESA and derived mass transfer stability thresholds for different kinds of stars \citep{2003MNRAS.341..662C,2008MNRAS.387.1416C,2014ApJ...796...37S,2015MNRAS.449.4415P}.

\begin{figure}
	\includegraphics[scale=0.45]{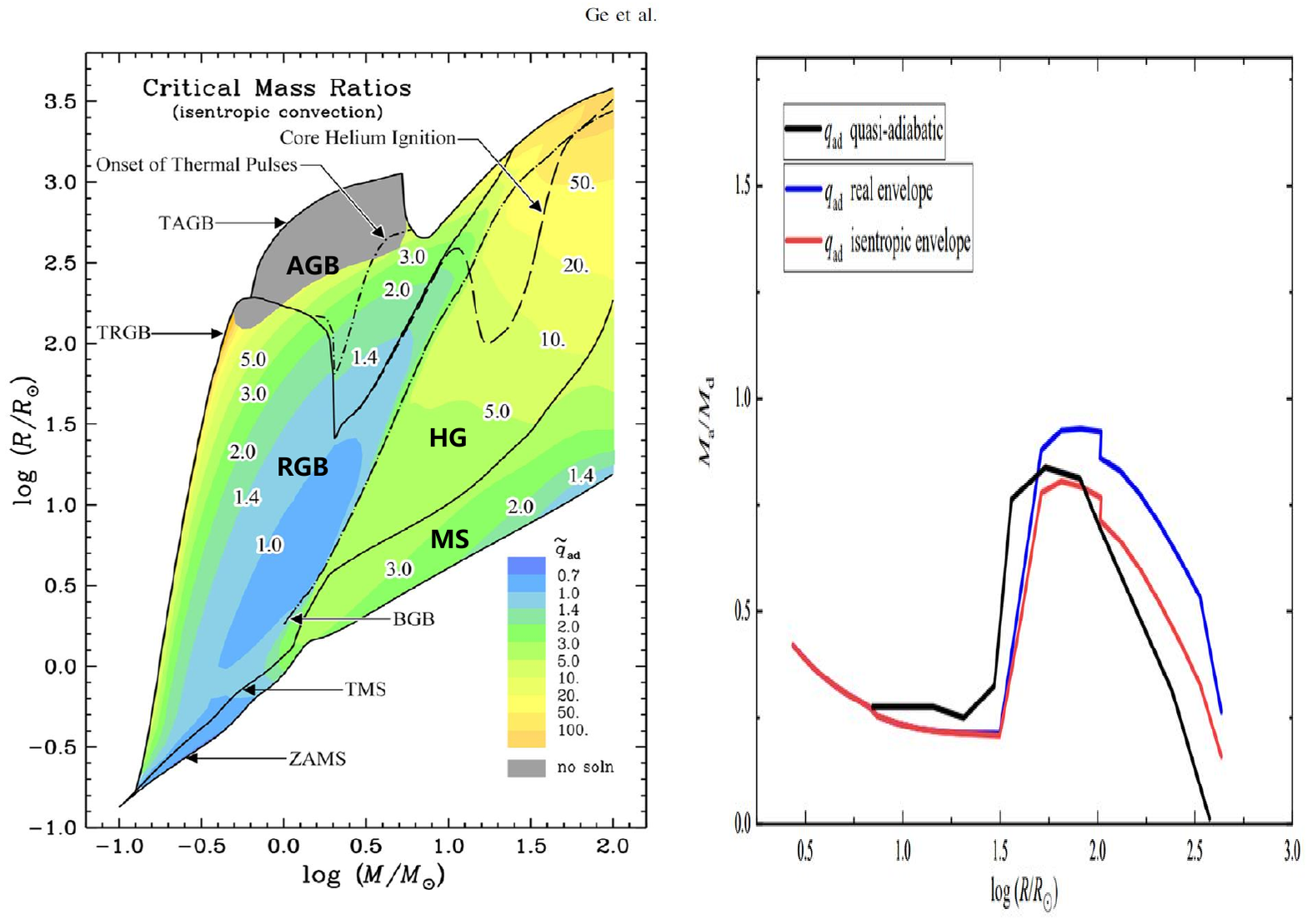}
	\caption{The critical mass ratios $q_{\rm crit}$ of dynamical timescale mass transfer for donor stars with different mass and evolution states (left panel) from \cite{2020ApJ...899..132G} and comparing of a 5 $M_\odot$ star's $q_{\rm crit}$ (right panel) between color lines from \cite{2020ApJ...899..132G} and the black line from \cite{2023A&A...669A..45T}.}
	\label{fig03}
\end{figure}

In the series of works by \citet{2010ApJ...717..724G,2010Ap&SS.329..243G,2015ApJ...812...40G,2020ApJ...899..132G}, Ge et al. built the adiabatic mass loss model for realistic instead of the polytropic equation of state stars ($Z=0.02$). These studies help overcom the non-realistic effects in evolved stars with partial ionization envelopes. The criteria of unstable mass transfer in these studies cover a full mass and evolutional state for potential donor stars (left panel in Figure\,\ref{fig03}). The abbreviations of the main sequence, Hertzsprung gap, red giant branch, and asymptotic giant branch are MS, HG, RGB, and AGB. They further extend their research to metal-poor stars \citep{2023ApJ...945....7G}, non-conservative mass transfer \citep{2024arXiv240816350G}, and helium-donor stars \citep{2024ApJS..274...11Z}. The critical initial mass ratios $q_{\rm crit} = M_{\rm donor}/M_{\rm accretor}$ for dynamical timescale mass transfer nicely predict the upper limits of observed mass ratios of cataclysmic variables \citep{2015ApJ...812...40G}, intermediate \citep{2023ApJ...945....7G} and high-mass \citep{2024arXiv240816350G} X-ray binaries. In addition to the comparison with observation, \citet{2023A&A...669A..45T} theoretically confirmed the results of low and intermediate-mass stars from \citet{2020ApJ...899..132G}.

Compared with the results of critical mass ratios from polytropic models, the major differences are for massive donor stars and RGB/AGB donor stars (left panel in Figure\,\ref{fig03}). For evolved massive MS/HG stars, the mass transfer tends to be more stable and highly related to the evolution states rather than a constant $q_{\rm crit} = 3$ or $4$. This result predicts more massive binary stars as GW sources of double BHs' or neutron-stars' (NSs) progenitors  might evolve through the stable mass transfer channel. For intermediate and low-mass RGB/AGB stars, the mass transfer also tends to be more stable (left panel in Figure\,\ref{fig03}) than the results of polytropic models. The unstable mass transfer for late RGB/AGB stars with $q_{\rm crit} \gtrsim 3$ might be dominated by a thermal timescale (shorter than 100 yrs) process through outer Lagrangian points \citep{2020ApJS..249....9G} instead of a prompt dynamical timescale case. These results can significantly impact the formation and evolution of double white dwarfs that are essential space GW sources of LISA \citep{2023LRR....26....2A} and Tianqin \citep{2024arXiv240919665L} projects.

\section{Application in double BHs/NSs}

Ever since the first direct detection of double smBHs merger by LIGO \citep{2016PhRvL.116f1102A}, about a hundred merging double compact objects (DCOs), i.e., BH+BH/BH+NS/NS+NS, have been detected \citep{2023PhRvX..13d1039A}. Among them, the $150\,M_\odot$ massive smBH merger \citep{2020PhRvL.125j1102A} and the large mass ratio DCO system \citep{2020ApJ...896L..44A} challeged the current knowledge about single and binary star evolution. However, these observations, in turn, bring single and binary stellar physics into a new development era.

The theoretical formation channels for DCOs include isolated binary evolution, for instance, a stable mass transfer (initial mass ratio $q_{\rm i} < q_{\rm crit}$), a CEE \citep{1976IAUS...73...75P} or a chemically homogeneous evolution (two components with similar masses) channel \citep{2016A&A...588A..50M,2016MNRAS.458.2634M}, and dynamical interactions in dense environments such as globular \citep{2000ApJ...528L..17P} and young stellar clusters \citep{2016MNRAS.459.3432M}. Among the isolate binary evolution channels, the CEE is initially considered the dominant formation channel for DCOs \citep{2002ApJ...572..407B}.

A recent systematic study indicates that mass transfer in massive binary systems \citep{2015ApJ...812...40G}, i.e., double BH progenitors, is more stable than previously found. After adopting the updated mass transfer thresholds by \citet{2015ApJ...812...40G}, the binary population synthesis study shows that the isolated binary BH formation scenario consisting of a stable mass transfer during the second mass transfer phase (instead of CEE) may dominate the double BH formation channel \citep{2019MNRAS.490.3740N}. Since then, together with the updated mass stability thresholds \citep{2015ApJ...812...40G,2020ApJ...899..132G,2017MNRAS.465.2092P,2021A&A...650A.107M,2021ApJ...920...81S} and the knowledge about the non-conserved mass transfer leading to an orbit shrinkage \citep[e.g.,][]{2006csxs.book..623T,2023ApJ...958..138W}, the stable mass transfer channel for DCOs becomes important in the GW community \citep{2021ApJ...922..110G,2021A&A...650A.107M,2021A&A...651A.100O,2021ApJ...920...81S,2022ApJ...931...17V,2022ApJ...940..184V,2023MNRAS.520.5724B,2024A&A...681A..31P,2024A&A...689A.305O}. Especially, \citet{2024A&A...681A..31P} use results from \citet{2010ApJ...717..724G,2015ApJ...812...40G,2020ApJ...899..132G} and resolve the parameter space of critical mass ratio for CEE to non-conservative mass transfer cases (slightly different with the method by \citealt{2024arXiv240816350G}). They find that the non-conservative stable mass transfer channel can produce merging binary BHs with large mass ratios, consistent with the majority of inferred sources of the third GW transient catalogue. Besides the mass transfer physics, the supernova natal kicks \citep[e.g.,][]{2021MNRAS.500.1380M,2021ApJ...920L..37W} and the rotation or extra-mixing \citep{2003A&A...404..975M} are also critical to be addressed in massive binary evolution. 

\section{Application in DWDs}

\begin{figure}
	\includegraphics[scale=0.45]{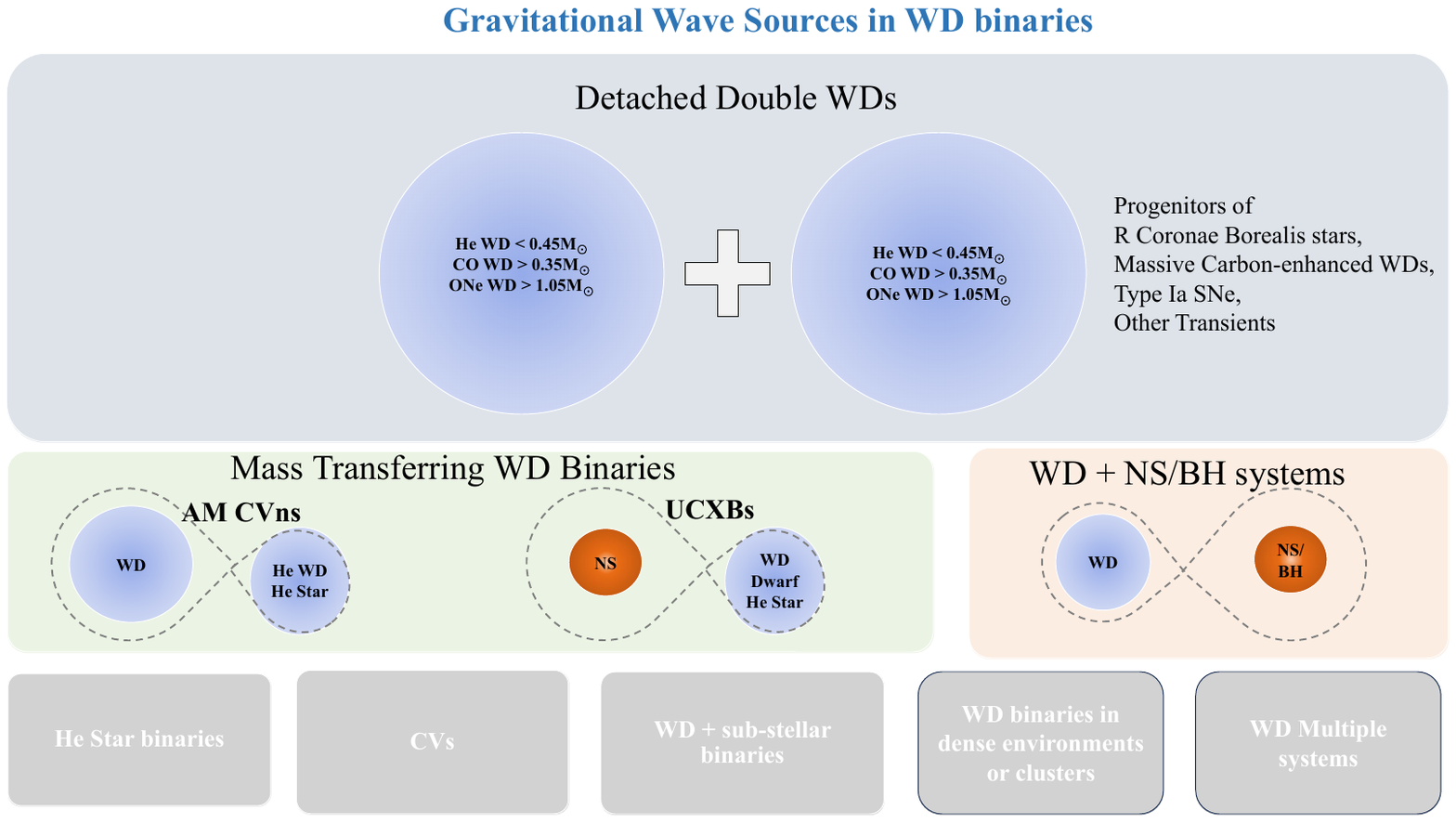}
	\caption{GW sources in many kinds of WD binaries adopted from \citet{2024arXiv240919665L}.}
	\label{fig04}
\end{figure}

Figure\,\ref{fig04} shows WD binaries, including DWDs, are essential GW sources for space LISA \citep{2023LRR....26....2A}/Tianqin \citep{2024arXiv240919665L} detectors. Combining GW detection and electromagnetic (EM) wave observation should significantly improve our knowledge of WD binaries and the binary evolution theory. The time-domain, spectroscopic observation, and the SDSS/Gaia survey have abundantly enlarged the observed WD binaries, like WD+main sequence (MS) stars \citep{2020ApJ...905...38R,2020MNRAS.494.3799P,2021MNRAS.506.2269E,2021MNRAS.506.5201R,2024MNRAS.527.8687O,2024PASP..136h4202Y}, and DWDs \citep{2018MNRAS.476.2584M,2020ApJ...889...49B,2021MNRAS.506.2269E}. These WD binaries are crucial sources for studying two fundamental problems, mass transfer stability and CEE in binary star evolution. DWDs are also essential sources for predicting the GW background for LISA/Tianqin.

Using the updated mass transfer stability thresholds from \citet{2010ApJ...717..724G,2015ApJ...812...40G,2020ApJ...899..132G} including critical mass ratios of dynamical timescale mass transfer for non-conservative cases later published by \citet{2024arXiv240816350G}, \citet{2023A&A...669A..82L} perform a series of binary population synthesis studies. \citet{2023A&A...669A..82L} find that for Ge et al.'s model, most of the DWDs are produced from the first stable non-conservative Roche lobe overflow plus the second
CE ejection channel instead of the first CE plus the second CE channel. \citet{2023A&A...669A..82L} show that the space densities for the detectable DWDs and those with extremely low-mass WD companions predicted by using Ge et al.'s model are close to what has been shown in observation. \citet{2023A&A...669A..82L} also claim the findings
from Ge et al.'s model predict a Galactic SN Ia rate that perfectly supports the observations. Finally, Ge et al.'s results support the observational DWDs merger rate distribution per Galaxy and the space density of double white dwarfs in the Galaxy \citep{2023A&A...669A..82L}.

\section{Acknowledgement}
We acknowledge support from the NSFC (grant Nos. 12288102, 12090040/3, 12125303, 12173081), the National Key R\&D Program of China (grant No. 2021YFA1600403/1), the key research program of frontier sciences, CAS, No. ZDBS-LY-7005, and Yunnan Fundamental Research Projects (grant NOs. 202101AV070001).

\end{document}